# Microwave-induced zero-resistance states in a high-mobility two-subband electron system

A. A. Bykov [1, 2], A. V. Goran [1], A. K. Bakarov [1]

[1] Rzhanov Institute of Semiconductor Physics, Novosibirsk, 630090 Russia
[2] Novosibirsk State University, Novosibirsk, 630090 Russia

In this study we used selectively-doped GaAs/AlAs heterostructure to fabricate a high-mobility two-subband electronic system with substantially different concentration of electrons in subbands. We observe microwave photoresistance at high numbers of magneto-intersubband oscillations (MISO). The system under study demonstrates microwave-induced resistance oscillations (MIRO) and MISO interference. MIRO in the studied two-subband system appear in lower magnetic fields comparing to MISO. This is an indication of some unknown mechanism that exists in the two-subband system and is responsible for MISO amplitude damping in low magnetic fields, while it does not affect the MIRO amplitude. Zero resistance states (ZRS) appear in the system under study under microwave irradiation in the narrow range of magnetic fields near the MISO maximum.

The discovery of microwave-induced resistance oscillations (MIRO) in high-mobility two-dimensional (2D) electron systems at high filling factors [1, 2] and zero resistance states (ZRS) [3, 4] appearing in the MIRO minimum inspired intensive theoretical and experimental research of these phenomena [5]. The search for new electron systems where these phenomena manifest themselves continues [6-8], as well as development of new theoretical models for their interpretation [9].

MIRO are periodic in the inverse magnetic field, similar to Shubnikov-de-Haas (SdH) oscillations. Their period is controlled by $\omega/\omega_c$ ratio and their minimum are placed at $\omega/\omega_c = j + \frac{1}{4}$, where $\omega = 2\pi f$ is microwave radiation frequency, $\omega_c = eB/m^*$ is cyclotron frequency of electrons with effective mass $m^*$ in magnetic field $B$, and $j$ is integer. The resistance of a 2D electron system exposed to microwave irradiation can be written as $\rho_\omega = \rho_{xx} + \rho_{MIRO}$, where $\rho_{xx}$ is the resistance in the absence of irradiation and $\rho_{MIRO}(\omega/\omega_c)$ is sign-alternating photoresistance.

There are two commonly used models that explain MIRO based on elastic and inelastic scattering of non-equilibrium electrons [10-13]. Both models produce the same expression for MIRO amplitude in the case of overlapping Landau levels [5]:

$$\rho_{MIRO} \propto - (\omega/\omega_c) P_\omega \lambda^2 \sin(2\pi\omega/\omega_c), \qquad (1)$$

where $P_\omega$ is a dimensionless parameter proportional to microwave power, $\lambda = \exp(-\pi/\omega_c\tau_q)$ is Dingle factor and $\tau_q$ is the quantum relaxation lifetime. According to (1) $\rho_\omega$ in the MIRO minimum should turn negative with increasing $P_\omega$. However the experiment shows $\rho_\omega \approx 0$ at the MIRO minimum [3, 4, 12, 14, 15]. It is commonly assumed that ZRS are the result of instability of absolute negative resistance which leads to formation of current domains [16-18].

Unlike single-subband electron systems, high-mobility two-subband electron systems exhibit MISO in perpendicular magnetic field in addition to SdH oscillations at low temperatures [19-21]. These quantum oscillations exist because of intersubband scattering and their period in the inverse magnetic field is $\Delta_{12}/\hbar\omega_c$, where $\Delta_{12}$ is the energy gap between subbands. MISO amplitude in two-subband system is [20, 21]:

$$\rho_{MISO} \propto \lambda_1 \lambda_1 \cos(2\pi\Delta_{12}/\omega_c), \qquad (2)$$

where $\lambda_1 = \exp(-\pi/\omega_c\tau_{q1})$, $\lambda_2 = \exp(-\pi/\omega_c\tau_{q2})$ and $\tau_{q1}$, $\tau_{q2}$ are quantum electron lifetimes in subbands.

Researches showed that microwave photoresistance in two-subband electron system exhibits MIRO-cut MISO [22]. MISO amplitude increases in the areas of positive photoresistance, while in the areas of negative photoresistance $\rho_{MIRO}$ inverts. This behavior was explained by MIRO and MISO interference [23]. ZRS in two-subband systems so far were only observed in a 45 nm width quantum well with close electron concentrations in subbands [24] where the condition $2\hbar\omega \sim \Delta_{12}$ was met. In this paper we observe ZRS under condition $2\hbar\omega \ll \Delta_{12}$, which allows far more detailed investigation of this non-equilibrium phenomenon in two-subband systems.

The selectively doped heterostructure under study was a single 30 nm width GaAs quantum well with side AlAs/GaAs superlattice barriers [25, 26]. Superlattice barriers were in the form of 15 alternating 1.41 nm width AlAs and 2.83 nm width GaAs layers. Charge carriers in the quantum well were provided by Si δ-doping. Single Si δ-doped layers were placed on both sides of the quantum well inside GaAs layers of superlattice barriers at the distance of 32.5 nm from the quantum well interfaces. The heterostructure was grown using molecular-beam epitaxy on (100) GaAs substrate.

The measurements were carried out at the temperature $T = 1.6$ K in the magnetic fields $B < 0.5$ T. Hall bars with the width $W = 50$ μm and the length $L = 450$μm were fabricated using optical photolithography and wet etching. The resistance of 2D electron gas was measured using alternating current of frequency 733 Hz, not exceeding $10^{-7}$ A. The concentration and mobility of electrons at $T = 1.6$ K were $n_H \approx 6.82\times10^{15}$ m$^{-2}$ and $\mu \approx 312$ m$^2$/Vs. The microwave irradiation was applied to the sample via a circular waveguide with an inner diameter of 6 mm. The sample was located at the distance of 1-2 mm from the open end of the waveguide. The microwave radiation power was $P_{out} \sim 4$ mW at the output of the generator.

Fig. 1a shows the dependence $\rho_{xx}(B)/\rho_0$ which reveals two types of oscillations. Only MISO exist at magnetic fields $B < 0.2$ T, while at $B > 0.2$ T they co-exists with SdH oscillations. Fourier analysis of the dependence $\rho_{xx}/\rho_0$ on $1/B$ is presented in Fig 1b. It reveals two frequencies corresponding to SdH oscillations ($f_{SdH1} \approx 9.87$ T, $f_{SdH2} \approx 4.09$ T) and one frequency corresponding to MISO ($f_{MISO} \approx 5.76$ T). Corresponding electron concentrations in subbands, calculated from the frequencies of oscillations, are $n_1 \approx 4.77\times10^{15}$ m$^{-2}$ and $n_2 \approx 1.98\times10^{15}$ m$^{-2}$. Hall concentration $n_H$ is a bit higher than the sum $n_1 + n_2 \approx 6.75\times10^{15}$ m$^{-2}$ which may be attributed to X electrons in the studied system [25, 26]. The difference $f_{SdH1} - f_{SdH2} \approx 5.78$ T matches MISO frequency, which confirms our interpretation of spectral dependency. Calculated from the MISO frequency, $\Delta_{12} \approx 9.83$ meV is in a good agreement with self-consistent calculation of zone structure of the studied GaAs quantum well.

Fig. 2a shows the dependence $\rho_{xx}(B)/\rho_0$ in the presence of microwave irradiation. It's clearly seen that microwave field substantially alters magnetoresistance of high-mobility two-subband electron system. The Fourier analysis of $\rho_\omega(1/B)/\rho_0$ is presented on Fig. 2b. It reveals four frequencies: $f_{MIRO}$, $f_{MISO} - f_{MIRO}$, $f_{MISO}$ and $f_{MISO} + f_{MIRO}$ with the highest peak corresponding to

MIRO. The period of these oscillations, similarly to single-subband systems [1, 2] is determined by $\omega/\omega_c$. Unlike single-subband systems the Fourier transform contains three frequencies: $f_{MISO}$ - $f_{MIRO}$, $f_{MISO}$ and $f_{MISO}$ + $f_{MIRO}$, which can be attributed to MIRO and MISO interference that was earlier discovered in two-subband systems with little difference in electron concentrations in subbands [22, 23]. However in the studied system the interference is clearly visible without frequency analysis due to the large difference between $f_{MIRO}$ and $f_{MISO}$.

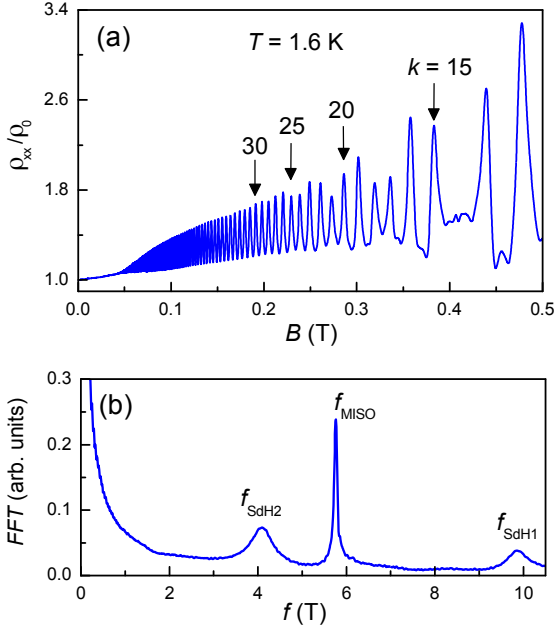

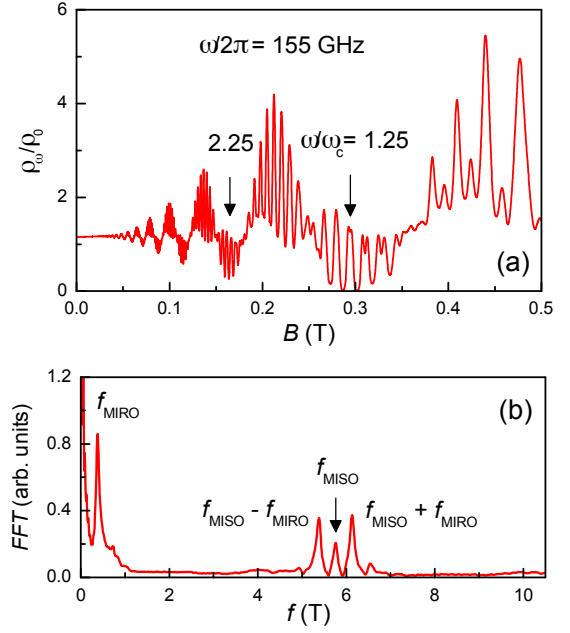

Fig. 1. (a) $\rho_{xx}(B)/\rho_0$ in two-subband electron system at $T = 1.6$ K. Arrows mark the MISO maximum numbered $k = 15, 20, 25$ and $30$ that correspond to $\Delta_{12} = k\hbar\omega_c$. (b) Fourier transform of $\rho_{xx}(1/B)/\rho_0$.

Fig. 2. $\rho_\omega(B)/\rho_0$ in two-subband electron system at $T = 1.6$ K in the presence of microwave irradiation of frequency $\omega/2\pi = 155$ GHz. Arrows mark the first and the second MIRO minimum, their position corresponds to $\omega/\omega_c = j + ¼$ for $j = 1$ and $2$. (b) Fourier transform of $\rho_\omega(1/B)/\rho_0$.

Fig. 3a shows dependences of $\rho_{xx}(B)/\rho_0$ and $\rho_\omega(B)/\rho_0$ at $B < 0.1$ T. As can be seen, MIRO manifest themselves at $B > 0.02$ T while MISO only appear at $B > 0.04$ T. The dependence of $\rho_{MIRO}(1/B)\omega_c/\rho_0\omega$ is linear in semilogarithmic scale (see Fig. 3b) which is in qualitative agreement with (1). The quantitative dependence of $\rho_{MIRO}(1/B)\omega_c/\rho_0\omega$ is:

$$\rho_{MIRO}\omega_c/\rho_0\omega = P_\omega \exp(-2\pi/\omega_c\tau_q^{MIRO}), \qquad (3)$$

where $P_\omega = 0.16$ and $\tau_q^{MIRO} = 15$ ps. It is still not clear whether $\tau_q^{MIRO}$ is determined by the quantum lifetimes $\tau_{q1}$ and $\tau_{q2}$ in the first and the second subband. The concentration $n_1$ is much higher than $n_2$ in the studied system so must be the contribution into $\rho_\omega$ from the first subband. In this case it seems natural to assume that $\tau_q^{MIRO} \approx \tau_{q1}$.

The experimental curve $\rho_{MISO}(1/B)/\rho_0$ is not linear in semilogarithmic scale (fig. 3b) and does not match (2). Such a behavior of MISO amplitude was earlier observed in the two-subband system with lower mobility [27]. It has been recently suggested that un-parabolic nature of electronic spectrum can be one of the possible reasons of MISO amplitude damping at low magnetic fields [28]. High scale fluctuation of the energy gap $\Delta_{12}$ caused by technological fluctuations of GaAs quantum well width could be another possible reason.

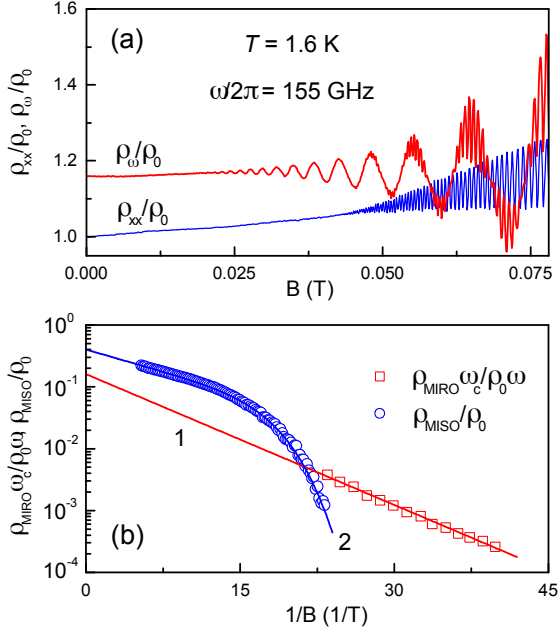 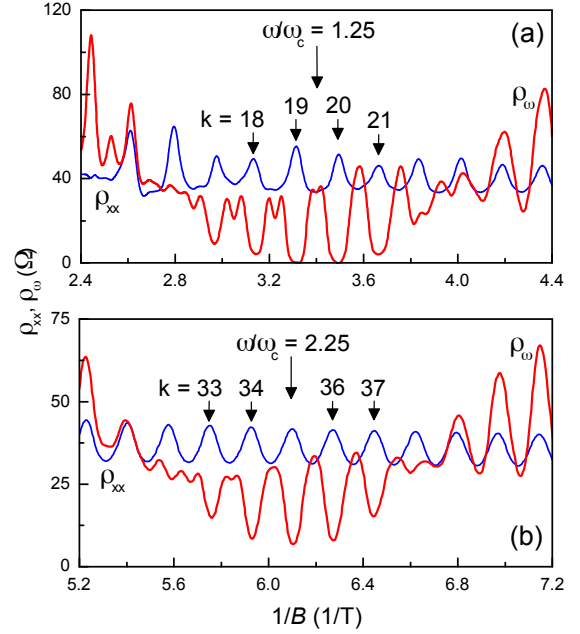

Fig. 3. (a) $\rho_{xx}(B)/\rho_0$ and $\rho_\omega(B)/\rho_0$ in two-subband electron system in low magnetic field. (b) $\rho_{MISO}(1/B)/\rho_0$ and $\rho_{MIRO}(1/B)\omega_c/\rho_0\omega$ in semilogarithmic scale. Circles and squares: experimental curves, solid lines: calculated curves. (1) is formula (3): $P_\omega = 0.16$, $\tau_q^{MIRO} = 15$ ps, (2) is formula (4): $A_{MISO} = 0.4$, $\tau_q^{MISO} = 22$ ps, $\tau_q^{DAMP} = 46$ ps, b = 6.

Fig. 4. Experimental curves $\rho_{xx}(1/B)$ and $\rho_\omega(1/B)$ for two-subband electron system at $T = 1.6$ K. Long arrows mark the positions of the first and the second MIRO minimum. Short arrows mark the MISO maximum.

In order to mathematically describe MIRO amplitude damping in low magnetic fields we introduced one more factor in addition to Dingle factor, similarly to how it was done for SdH oscillations amplitude in non-homogeneous 2D electron gas [29]:

$$\rho_{MISO}/\rho_0 = A_{MISO} \exp(-2\pi/\omega_c\tau_q^{MISO}) \exp(-2\pi/\omega_c\tau_q^{DAMP})^b, \qquad (4)$$

where $A_{MISO}$ is a dimensionless parameter, $2/\tau_q^{MISO} = 1/\tau_{q1} + 1/\tau_{q2}$, $\tau_q^{DAMP}$ is quantum time that takes into account MISO damping at low magnetic fields, $b$ is the power exponent. This equation quantitatively describes $\rho_{MISO}/\rho_0$ in the whole experimental range of magnetic fields with the following fitting parameters: $A_{MISO} = 0.4$, $\tau_q^{MISO} = 22$ ps, $\tau_q^{DAMP} = 46$ ps, b = 6.

The comparison of experimental curves $\rho_{MIRO}(1/B)\omega_c/\rho_0\omega$ and $\rho_{MISO}/\rho_0(1/B)$ leads us to a conclusion that the unknown mechanism responsible for MISO damping in low magnetic fields does not influence, or has very limited influence on the the MIRO amplitude. Assuming that $\tau_q^{MIRO} \approx \tau_{q1}$ we get the value of $\tau_{q2} \approx 41$ ps. In any case neither of times $\tau_{q1}$ and $\tau_{q2}$ can be lower than $\tau_q^{MIRO} \approx 15$ ps. States with $\rho_\omega \approx 0$ in the two-subband system with with $n_1 \approx n_2$ and $\tau_{q1} \approx \tau_{q2}$ were observed with $\tau_q \approx 7.1$ ps [24]. The high quality of the two-subband system under study can be confirmed by the dependence shown in Fig. 4a. It's clearly seen that MISO inverts around the first MIRO minimum ($\omega/\omega_c = 1 + 1/4$) and oscillation maximum numbered $k$=19 and 20 transform to the states with $\rho_\omega \approx 0$. However the state with $\rho_\omega \approx 0$ does not appear around the second MIRO minimum (fig. 4b) in the studied system.

In conclusion, in this paper we studied magnetotransport properties of high-mobility two-subband electron system ($\mu > 300$ m$^2$/Vs at $T = 1.6$ K) fabricated using selectively doped 20 nm width GaAs quantum well with side AlAs/GaAs superlattice barriers. It was shown that MIRO and ZRS appear in such a system under microwave irradiation. ZRS only appear at the MIRO minimum in the narrow range of magnetic fields near MISO maximum, which is attributed to MIRO and MISO interference. The unknown mechanism exists in the studied system that is responsible for MISO damping in low magnetic fields ($B < 0.1$ T) and does not affect MIRO amplitude. The analysis of MIRO amplitude dependence on $1/B$ at $B < 0.05$ T showed that the electron quantum lifetime in subbands exceeds 15 ps at $T = 1.6$ K in our samples. We believe that two-subband systems with high electron quantum lifetime can be uses for narrow-band receiving of microwave radiation.

This work was performed with support from RFFI (project 16-02-00592).